# Impact of the State of Emergency Declaration for COVID-19 on Preventive Behaviors and Mental Conditions in Japan: Difference in Difference Analysis using Panel Data

EIJI YAMAMURA* AND YOSHIRO TSUTSUI

May 2020 [1]

*Abstract*

*During the COVID-19 epidemic in Japan between March and April 2020, Internet surveys were conducted to construct panel data to investigate changes at the individual level regarding preventive behaviors and mental conditions by surveying the same respondents at different times. Specifically, the difference-in-difference (DID) method was used to explore the impact of the COVID-19 state of emergency declared by the government. Key findings were: (1) the declaration led people to stay home, while also generating anger, fear, and anxiety. (2) The effect of the declaration on the promotion of preventive behaviors was larger than the detrimental effect on mental conditions. (3) Overall, the effect on women was larger than that on men.*

---

[1] Corresponding author: Eiji Yamamura, Seinan Gakuin University, 6-2-92 Nishijin Sawaraku Fukuoka, 814-8511, Japan (e-mail:yamaei@seinan-gu.ac.jp). Coauthor: Yoshiro Tsutsui, Kyoto Bunkyo University, Japan (email: tsutsui@econ.osaka-u.ac.jp);. Acknowledgments: We would like to thank Editage [http://www.editage.com] for editing and reviewing this manuscript for English language.

# I. Introduction

The COVID-19 epidemic has had a significant impact on social and economic conditions, resulting in drastic changes in lifestyle, even though only a few months have passed since the first infected person was found in China in November 2019. Policymakers have been implementing various measures to mitigate COVID-19 pandemic. On May 16, 2020, the USA's death toll rose to 85,813, making it the highest official figure in the world, which was almost 2.5 times larger than UK, French, Italy, and Spain. At the same time, Japan's death toll was only 687[2]. A question arises here; was government's policies for COVID-19 control more effective than the USA and other countries? In this note, we examined the question in setting a quasi-natural experiment.

In various countries including China, UK, and the USA, governments have employed a strong measure of lockdown enforcement in response to surging numbers of persons infected by the virus. Countries that have implemented drastic measures, such as lockdown, have seen reductions in the speed of the pandemic spread (Fang, Wang, & Yang, 2020; Tian et al., 2020). On the other hand, a rapid increase in domestic violence has been observed (WHO 2020). However, the negative effects of government policies have not been not sufficiently investigated, as even the closure of schools and non-essential businesses, could increase psychological costs such as the deterioration of mental conditions.

Fetzer et al. (2020) conducted a large-scale web survey between late March and early April covering 58 countries to investigate preventive behaviors and mental conditions

---

[2] Johns Hopkins University of Medicine, CORONAVIRUS RESOURCE CENTER. https://coronavirus.jhu.edu/map.html. (Accessed on May 16, 2020)

within the population. Furthermore, they assessed the changes in the evaluations that people made during this period concerning government policies. However, they did not survey the same respondents to construct panel data and the sample was comprised of countries with different economic and social conditions. Therefore, they could not disentangle the effects of government policies on people's perceptions and behaviors from those other factors. Layard et al. (2020) analyzed the costs and benefits of the lockdown in the UK by considering not only traditional economic indices such as income and unemployment but also mental health [3]. However, no study has evaluated the effectiveness of policy measures to mitigate the COVID-19 epidemic by random assignment, although a few studies have assessed this aspect using simulation [4]. "One possible reason for the lack of systematic testing of mitigation measures in the COVID-19 pandemic is that it is both ethically and practically challenging: standard impact evaluation approaches typically require random assignment of some regions to an intervention and others to a control condition" (Haushofery and Metcalf, 2020, p.3).

On April 7, 2020, the Japanese government declared a state of emergency for COVID-19 in the prefectures where the number of persons infected with the virus was very large. Nine prefectures were clearly affected by the COVID-19 epidemic. However, the declaration was held only for seven of these nine prefectures. In this study, the seven prefectures where the declaration was held were defined as the treatment group. The two remaining prefectures were defined as the control group. The number of individuals infected by COVID-19 in the treatment group was not statistically different from that of the control group, as shown in Figure 1. On the other hand, there was a remarkable and statistically significant difference between the treatment and the control groups. The reason why the prefectures of Hokkaido and Aichi, which were the control group, were excluded from the declaration is ambiguous. This type of setting can be considered as quasi-randomization.

---

[3] Oswald and Pawdthavee (2020) indicated that releasing from lockdown UK citizens aged 20-30 years who did not live with their parents was effective in increasing economic and social benefits and did not lead to higher numbers of COVID-19 victims.

[4] The Susceptible-Infected-Recovered (SIR) model is used to compare the time course of infections in hypothetical control and treatment groups (Alvarez et al., 2020; Haushofery and Metcalf 2020). Atkeson (2020) also used the SIR model to consider the time course by dividing the total population into groups susceptible to the disease, infected by COVID-19, and others.

[Insert Figure 1 Here]

In Japan, similarly to the USA and European countries, governments have asked the population to change hygiene and social behaviors to help contain the spread of the disease (e.g., washing hands more carefully and avoiding social gatherings). Later, a request for more strict and costly measures, such as school closure and staying at home, was implemented. Differing from lockdown enforced in other countries, such as Italy, France, Germany, the UK, and the USA, the declaration of a state of emergency by the Japanese Government could not substantially penalize by law those individuals who did not obey the government's request. In other words, citizens in Japan could decide whether or not to carry out preventive behaviors to mitigate the COVID-19 pandemic. Therefore, it seems plausible that the reactions of Japanese citizens to COVID-19 may vary compared to countries imposing more rigorous lockdowns. The situation in Japan can be considered a natural experiment to examine how the government's policy to appeal to conscience and morale generates changes in citizens' behaviors, which in turn affect their mental conditions under a state of emergency [5].

The contribution of this study is to assess changes in preventive behaviors and mental conditions after a short period under the COVID-19 pandemic. For this purpose, the treatment and control groups were compared in order to explore how policies requiring preventive behaviors without penalty were effective in setting a quasi-natural experiment.

## II. Setting and overview of data

A. *Setting*

In Japan, the first person infected by COVID-19 was observed on January 16, 2020. The number of infected persons has increased as time has passed, however, the pace of increase is much slower than that of the USA. We carried out Internet surveys to gather data concerning citizens' preventive behaviors regarding COVID-19 and their mental conditions, exploring how they behaved and felt in response to the emergent situation of the pandemic.

---

[5] Ito et al. (2018) found that moral suasion was useful to persuade Japanese citizens to follow the request of saving electricity in a short period.

INTAGE, a company with significant experience in academic research, was commissioned to conduct the Internet survey. The sampling method used was designed to gather a representative sample of the Japanese population regarding gender, age, and prefecture of residence. Our survey selected Japanese citizens between 16 and 79 years old from all regions of Japan. We conducted online surveys three times to assess the same individuals and construct the panel data within a month.

In the first wave, the sample size was 4,359, and its response rate was 54.7 %. In the second and third waves, we surveyed respondents from the first wave, and their response rates were 80.2% and 92.2%, respectively. Totally, observations were 11,867, which included 4,359 individuals. The first wave of the survey was conducted between March 13 and 16. The second and the third waves were carried out between March 27 and 30 and between April 10 and 13, respectively. During the whole period studied, the number of infected individuals in Japan increased from 675 (first wave) to 1,387 (second wave), and then 5,347 (third wave).

On April 7, between the second and third waves, the Government of Japan declared a state of emergency for seven prefectures that had heavily suffered from COVID-19, including Tokyo and Osaka[6]. The declaration requested that people should avoid going out of home unnecessarily, and also requested closing various public places including schools, museums, theaters, and bars, among others. At the time of the declaration, the request was planned to be valid for one month. Therefore, immediately after the declaration, the third wave of the survey was carried out. We were able to examine the effects of the declaration on citizens' behaviors and mental conditions by comparing them before and after the declaration.

---

[6] Besides Tokyo and Osaka, the following prefectures were included in the survey: Kanagawa, Chiba, Saitama, Hyogo, and Fukuoka.

The novel setting for this study was that the state of emergency was not declared in two prefectures, Hokkaido and Aichi, even though the number of infected persons in these prefectures was almost equivalent to that of the other seven prefectures. We divided the 47 prefectures into the treatment group, comprised of the seven prefectures which were the target of the declaration, the control group composed of the prefectures of Hokkaido and Aichi, and other prefectures. Figure 1 compares the mean values of individuals infected with COVID-19 between groups, showing no statistically significant differences between the treatment and control groups, while significant differences were found between the control and treatment groups when compared to other groups. Therefore, citizens from the treatment group and those from the control group experienced almost the same situation.

The population of the nine prefectures covering the treatment and control groups was equivalent to 53.1% of the population of Japan. Specifically, the seven prefectures of the treatment group signified 43.1% of the Japanese population. In the sample used in this survey, eight prefectures had megacities of over one million people, [7] with the exception of

---

[7] Chiba prefecture included Chiba city which has 0.97 million people.

Chiba prefecture.

B. *Features of the data*

Observations for the treatment and control groups were 5,492 and 1,269, respectively. Therefore, total observations were 6,761. The survey questionnaire contained basic questions about demographics such as age, gender, educational background, household income, job status, marital status, and number of children. These data were constant in the first, second and third waves because the three waves were conducted within a month. In addition, respondents were asked questions concerning preventive behaviors, which are mentioned as follows:

*"Within a week, to what degree have you achieved the following behaviors? Please answer in a scale from 1 (I have not achieved this behavior at all) to 5 (I have completely achieved this behavior)."*

(1) Stay indoors

(2) Not go to the workplace (or school)

(3) Wearing a mask

(4) Washing hands carefully

The answers to these questions were proxies for preventive behaviors: *Stay indoors, Do not work, Wear mask,* and *Washing hands.*

With regards to mental conditions, respondents were asked the following question:

*"How much have you felt the emotions of anger, fear, and anxiety? Please answer in a scale from 1 (I have not felt this emotion at all) to 5 (I have felt this emotion strongly)."*

(5) Anger

(6) Fear

(7) Anxiety

The answers to these questions were proxies for mental conditions: *Anger, Anxiety,* and *Fear.*

[Insert Figure 2 Here]

[Insert Figure 3 Here]

Figures 2 and 3 demonstrate the degree of changes in preventive behaviors and mental conditions of respondents between March 13 and April 10 by comparing the treatment and control groups. In the next section, we will explain the difference-in-difference (DID) method to examine the effects of the declaration. Based on this method, there was the key assumption that the trends of variables would be the same for the treatment and control groups before the declaration was announced (Angrist and Pschke, 2009). In Figures 2 and 3, we examined the common trends assumption, confirming that the trends regarding preventive behaviors and mental conditions were almost the same in the control and treatment groups, between March 13 and 27, before the declaration of a state of emergency. Therefore, the common trends assumption was confirmed in this study.

Between March 13 and 27, except for "washing hands," levels of variables in the treatment group were lower than those of the control group. Later, between March 27 and April 10, the slopes of the treatment group became steeper than those of the control group, leading the mean values of the treatment group to be higher than those of the control group, with the exception of "washing hands." This suggests that the declaration had a significant effect on citizens' behaviors and mental conditions.

### III. The econometric model

The DID method was used to examine the effects of the state of emergency declaration on preventive behaviors and mental conditions. Data was limited to nine prefectures and divided into the control and treatment groups, which included residents from two and seven prefectures, respectively. As discussed in the previous section, the DID method was considered valid. The estimated function takes the following form:

$$Y_{itp} = a_1\ Wave3_t \times Treatment_g + a_2\ Wave3_t + a_3\ Wave2_t + a_4\ Infected\ COVID19_{itg} + k_i + u_{itp},$$

In this formula, $Y_{itp}$ represents the dependent variable for individual $i$, wave $t$, and group $g$. For the estimation of preventive behaviors, *Y* is preventive behaviors such as *Stay indoors, Do not work, Wear mask,* and *Washing hands.* Regarding the estimation of mental conditions, *Y* is *Anger, Anxiety,* and *Fear.* The second (*Wave 2*) and third wave (*Wave 3*) dummies were included while their reference group was the first wave. This describes the degree of change in the dependent variables compared to the first wave. *Treatment* is a dummy variable for the treatment group. Key variable was, *Wave3* ×*Treatment*, cross terms for *Wave3* and *Treatment.* When preventive behaviors were dependent variables, their coefficients were expected to be positive if the declaration promoted citizens to achieve preventive behaviors. On the other hand, when mental conditions were dependent variables, their coefficients were expected to be positive if the declaration deteriorated the mental conditions of the population.

The time-invariant individual-level fixed effects are represented by $k_i$. Because of short-term panel data, most of the individual level demographic variables were considered as time-invariant features, which were completely described by $k_i$. The regression parameters are denoted by $a$. The number of persons infected with COVID-19 increased drastically in the residential areas during the studied period and, thus, it was included as a control variable. The error term was denoted by $u$.

**IV. Results**

Results are focused on the key variable *Wave3 $_t$ × Treatment*. The results shown in Table 1 are based on the sub-sample comprised of the treatment and control groups shown in Figure 1. In this table, cross terms display a positive coefficient in all results [8]. Furthermore, we observe a statistical significance at a level of 1 %, with the exception of *"Washing hands,"* which was not significant. The appropriate setting of the control group shows strong evidence that declaring a state of emergency promoted preventive behaviors while at the same time deteriorated mental health. With regards to preventive behaviors, the coefficients for *"Stay indoors"* and *"Do not work"* were larger than those for *"Wear mask"* and *"Washing hands."* Our interpretation of these results is that citizens were requested to wear a mask and wash hands mainly when they went out. Naturally, people may have considered staying indoors as more important than wearing a mask and washing hands.

[Insert Table 1 Here]

To investigate gender differences concerning the effects of the declaration, we conducted estimations using sub-samples for males and females. In Table 2, we observe a positive coefficient in all cross terms. Differences between males (Panel A) and females

---

[8] The Appendix presents results based on the full sample, where the control group was comprised of respondents from 40 prefectures including the "Control group" and "Others," indicated in Figure 1. Cross terms show a positive coefficient with the exception of "Washing hands." Furthermore, statistical significance was observed for "Stay indoors," "Do not work," and "Anger."

(Panel B) are shown below.

[Insert Table 2 Here]

First, all columns for females show a statistical significance, whereas statistical significance was not observed regarding two cross terms, "*Wash hands*" and "*Anxiety*." Hence, as a whole, the declaration had more significant effects on females than in males. Washing hands is different from other preventive behaviors in that it is a behavior less likely to be observed by other people. This seems to reduce the incentive to wash hands. In other words, the statistical significance observed in females could mean that women may have the incentive to wash their hands even if other people are not observing their behavior. As Aguero and Beleche (2017) indicate, an exogenous health shock, such as a pandemic, facilitates the adoption of low-cost health behaviors, such as hands washing, which provides long-lasting effects on health outcomes. Therefore, the role of women becomes important to have a long-term effect on the general acceptance of handwashing in a society.

Second, the coefficient value for "*Anger*" in males was two times larger than in females. We may interpret this as suggesting that an increase in anger in husbands could result in an increase in domestic violence against their wives during the state of emergency. On the other hand, females were more likely to feel anxiety and fear than males, which seems to result in mental illness. This could cause social agitation.

Overall, the declaration of a state of emergency not only had positive effects on preventive behaviors addressed to mitigate the pandemic, but also negative effects on mental conditions which may increase domestic violence and social unrest. However, in most of cases, the coefficient values of preventive behaviors were larger than those of mental conditions.

## V. Conclusion

The evaluation of government policies should be analyzed by considering their costs and benefits. The purpose of this study was to examine how the declaration of a state of emergency for COVID-19 changed preventive behaviors and mental conditions in Japan. Using individual-level panel data collected through short-term repeated Internet surveys, we conducted DID estimations. After controlling for individual fixed-effects, key findings were: (1) the declaration led people to stay home, while also generating anger, fear, and anxiety. (2) The effect of the declaration on the promotion of preventive behaviors was larger than the detrimental effect on mental conditions. (3) Overall, the effect on women was larger than that on men. In short, we found that the declaration promoted preventive behaviors and at the same time deteriorated mental conditions. More specifically, an increase in anger in husbands is remarkably larger than wives, which could result in an increase in domestic violence against their wives during the state of emergency.

An increase in anger from staying indoors is thought to cause domestic violence. Considering this aspect is important when evaluating the outcomes of the state of emergency declaration in Japan as well as lockdowns in Italy, France, Spain, the United Kingdom, and the United States. Moreover, it is necessary to evaluate government policies through cost-benefit analysis from a long-term viewpoint. Further research should investigate these aspects to scrutinize whether Japanese government's policy is more effective and efficient than policies adopted by the USA, UK, French, Italy, and Spain.

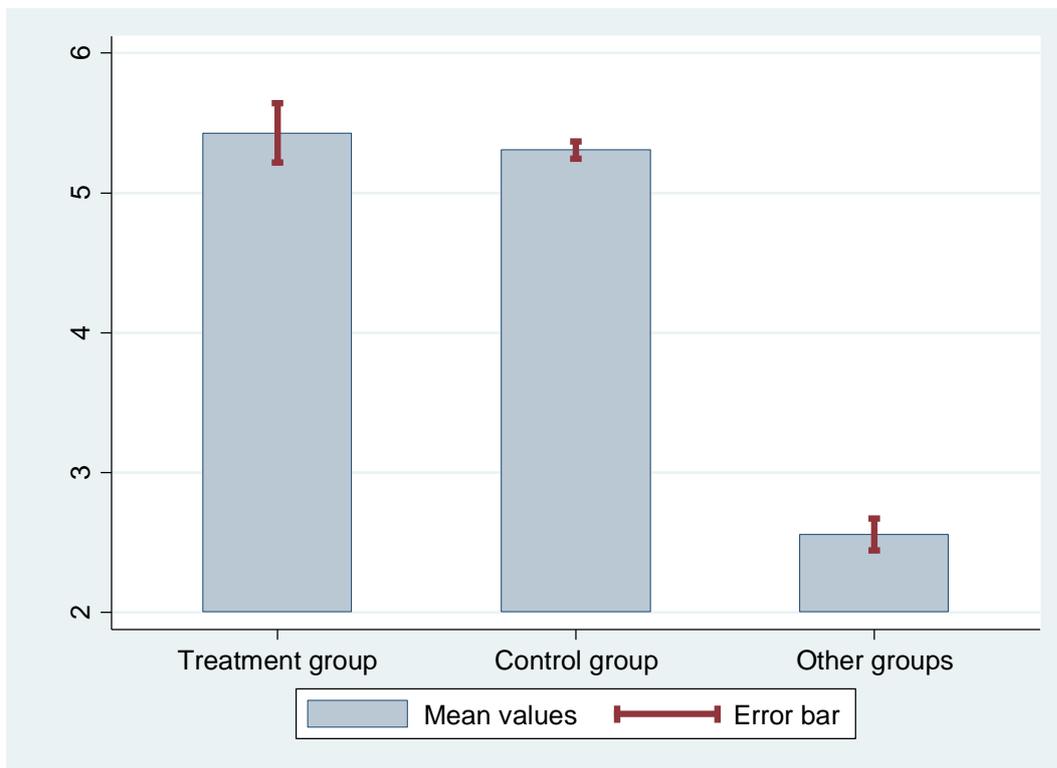

**Figure 1.** Comparison of mean number of persons infected by COVID-19 in each group between April 1 and 7, 2020.

**Note:** The error bar represents the 95% confidence intervals.

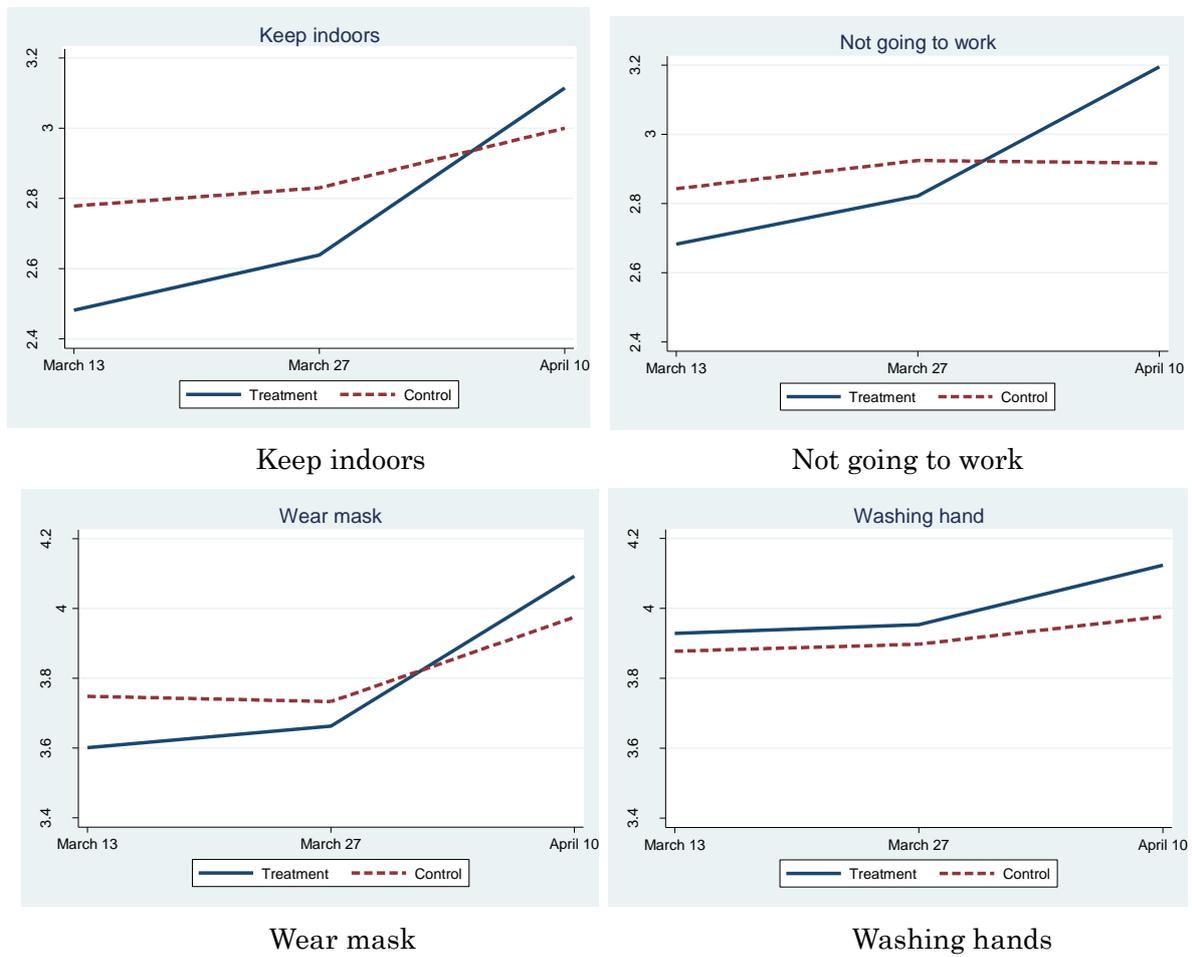

**Figure 2.** Changes in preventive behaviors.

Note: The solid line indicates the treatment group, while dashed line shows the control group.

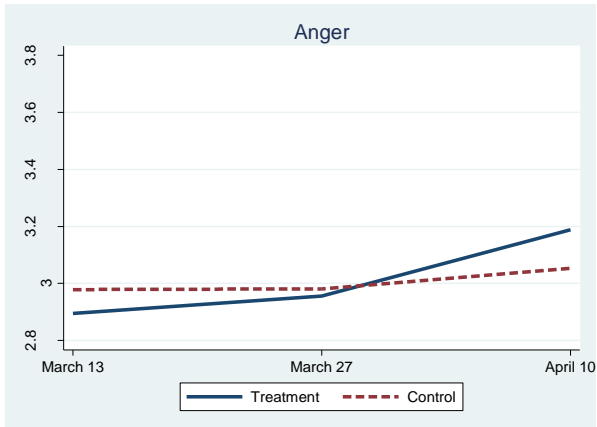
Anger

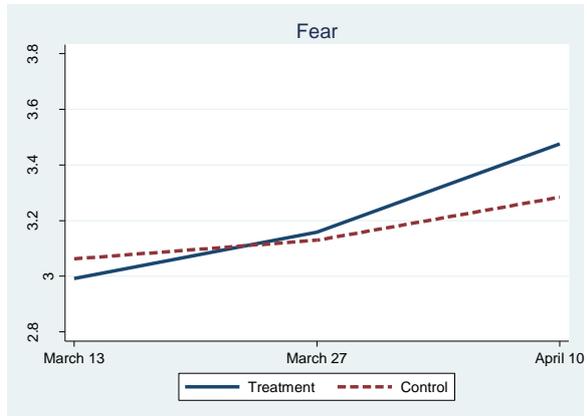
Fear

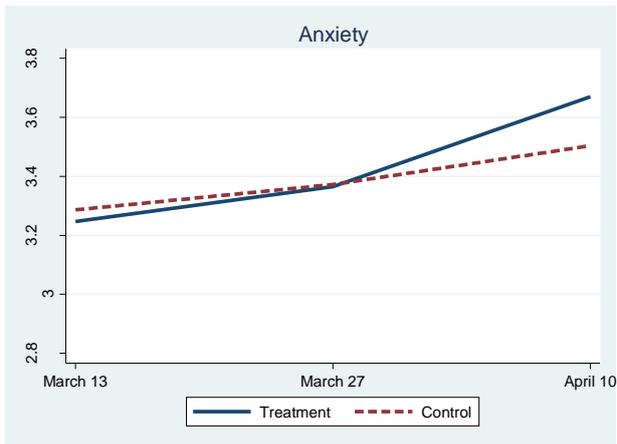
Anxiety

**Figure 3.** Changes in emotions.

Note: The solid line indicates the treatment group, while the dashed line shows the control group.

**Table 1.** Fixed effects model: Sub-sample of heavily infected areas.

|  | Preventive behaviors | | | | Mental conditions | | |
|---|---|---|---|---|---|---|---|
|  | *Keep indoors* | *Not going to work* | *Wear mask* | *Washing hands* | *Anger* | *Anxiety* | *Fear* |
| *Wave3 ×Treatment* | 0.29*** (0.06) | 0.32*** (0.07) | 0.19*** (0.06) | 0.06 (0.04) | 0.18*** (0.06) | 0.15*** (0.05) | 0.15*** (0.05) |
| *Wave3* | 0.21*** (0.05) | 0.07 (0.06) | 0.27*** (0.05) | 0.11*** (0.03) | 0.11** (0.05) | 0.28*** (0.05) | 0.24*** (0.05) |
| *Wave2* | 0.10*** (0.03) | 0.08** (0.03) | 0.07*** (0.02) | 0.03 (0.02) | 0.05** (0.02) | 0.15*** (0.02) | 0.11*** (0.02) |
| *Infected COVID_19* | 0.23 (0.07) | 0.23*** (0.07) | 0.05 (0.06) | 0.03 (0.04) | −0.02 (0.06) | 0.07 (0.05) | 0.05 (0.05) |
| Within R-Square | 0.11 | 0.06 | 0.09 | 0.02 | 0.03 | 0.08 | 0.07 |
| Groups | 2,477 | 2,477 | 2,477 | 2,477 | 2,477 | 2,477 | 2,477 |
| Obs. | 6,761 | 6,761 | 6,761 | 6,761 | 6,761 | 6,761 | 6,761 |

Note: Numbers within parentheses indicate robust standard errors clustered on individuals.

***, ***, * indicate statistical significance at a level of 1%, 5%, and 10%, respectively.

**Table 2.** Fixed effects model: Sub-sample.

Panel A. Male sample

|  | | Preventive behaviors | | | | Mental conditions | |
|---|---|---|---|---|---|---|---|
|  | *Keep indoors* | *Not going to work* | *Wear mask* | *Washing hands* | *Anger* | *Anxiety* | *Fear* |
| *Wave3 ×Treatment* | 0.32*** (0.09) | 0.21** (0.10) | 0.17* (0.09) | 0.002 (0.03) | 0.22** (0.09) | 0.10 (0.08) | 0.14* (0.08) |
| *Wave3* | 0.12 (0.08) | 0.08 (0.09) | 0.29*** (0.07) | 0.14** (0.06) | 0.05 (0.08) | 0.29*** (0.07) | 0.27*** (0.07) |
| *Wave2* | 0.05 (0.04) | 0.03 (0.04) | 0.05 (0.03) | 0.01 (0.03) | 0.05 (0.04) | 0.15*** (0.04) | 0.12*** (0.03) |
| *Infected COVID_19* | 0.21** (0.09) | 0.29*** (0.10) | 0.05 (0.09) | 0.04 (0.06) | 0.02 (0.09) | 0.11 (0.08) | 0.002 (0.08) |
| Within R-Square Groups Obs. | 0.09 1,206 3,296 | 0.05 1,206 3,296 | 0.09 1,206 3,296 | 0.02 1,206 3,296 | 0.03 1,206 3,296 | 0.08 1,206 3,296 | 0.09 1,206 3,296 |

Panel B. Female sample

|  | | Preventive behaviors | | | | Mental conditions | |
|---|---|---|---|---|---|---|---|
|  | *Keep indoors* | *Not going to work* | *Wear mask* | *Washing hands* | *Anger* | *Anxiety* | *Fear* |
| *Wave3 ×Treatment* | 0.25*** (0.09) | 0.43*** (0.10) | 0.22*** (0.08) | 0.11** (0.05) | 0.14* (0.08) | 0.20** (0.08) | 0.16** (0.08) |
| *Wave3* | 0.30*** (0.07) | 0.06 (0.08) | 0.25*** (0.06) | 0.09* (0.05) | 0.17** (0.07) | 0.27*** (0.07) | 0.21*** (0.07) |
| *Wave2* | 0.14*** (0.04) | 0.12*** (0.04) | 0.10*** (0.03) | 0.04 (0.03) | 0.06* (0.03) | 0.15*** (0.03) | 0.10*** (0.03) |
| *Infected COVID_19* | 0.25*** (0.09) | 0.18* (0.11) | 0.06 (0.08) | 0.02 (0.05) | −0.06 (0.07) | 0.04 (0.07) | 0.09 (0.07) |
| Within R-Square Groups Obs. | 0.14 1,271 3,465 | 0.07 1,271 3,465 | 0.10 1,271 3,465 | 0.03 1,271 3,465 | 0.03 1,271 3,465 | 0.09 1,271 3,465 | 0.08 1,271 3,465 |

Note: Numbers within parentheses indicate robust standard errors clustered on individuals.

***, ***, * indicate statistical significance at a level of 1%, 5%, and 10%, respectively.

**Appendix. Results using the full sample.**

|  | | Preventive behaviors | | | | Mental conditions | |
|---|---|---|---|---|---|---|---|
|  | *Keep indoors* | *Not going to work* | *Wear mask* | *Washing hands* | *Anger* | *Anxiety* | *Fear* |
| *Wave3 ×Treatment* | 0.11** (0.05) | 0.23*** (0.05) | 0.05 (0.04) | −0.02 (0.03) | 0.08** (0.04) | 0.06 (0.04) | 0.05 (0.05) |
| *Wave3* | 0.40*** (0.03) | 0.17*** (0.03) | 0.41*** (0.03) | 0.20*** (0.02) | 0.19*** (0.02) | 0.36*** (0.02) | 0.33*** (0.02) |
| *Wave2* | 0.13*** (0.02) | 0.09*** (0.02) | 0.08*** (0.02) | 0.04*** (0.01) | 0.03 (0.02) | 0.13*** (0.02) | 0.09*** (0.02) |
| *Infected COVID_19* | 0.21*** (0.07) | 0.23*** (0.07) | 0.04 (0.06) | 0.02 (0.04) | −0.02 (0.06) | 0.07 (0.05) | 0.05 (0.04) |
| Within R-Square | 0.09 | 0.04 | 0.09 | 0.03 | 0.03 | 0.07 | 0.07 |
| Groups | 4,359 | 4,359 | 4,359 | 4,359 | 4,359 | 4,359 | 4,359 |
| Obs. | 11,867 | 11,867 | 11,867 | 11,867 | 11,867 | 11,867 | 11,867 |